# Beat the Rayleigh limit: OAM based super-resolution diffraction tomography

Lianlin Li[1] and Fang Li[2]

[1]*Center of Advanced Electromagnetic Imaging, Department of Electronics, Peking University, Beijing 100871, China*

[2]*Institute of Electronics, CAS, Beijing 100190, China*

This letter is the first to report that a super-resolution imaging beyond the Rayleigh limit can be achieved by using classical diffraction tomography (DT) extended with orbital angular momentum (OAM), termed as OAM based diffraction tomography (OAM-DT). It is well accepted that the orbital angular momentum (OAM) provides additional electromagnetic degrees of freedom. This concept has been widely applied in science and technology. In this letter we revisit the DT problem extended with OAM, demonstrate theoretically and numerically that there is no physical limits on imaging resolution in principle by the OAM-DT. This super-resolution OAM-DT imaging paradigm, has no requirement for evanescent fields, subtle focusing lens, complicated post-processing, etc., thus provides a new approach to realize the wavefield imaging of universal objects with sub-wavelength resolution.

In 1879 Lord Rayleigh formulated a criterion that a conventional imaging system cannot achieve a resolution beyond that of the diffraction limit, i.e., the Rayleigh limit. Such a "physical barrier" characters the minimum separation of two adjacent objects that an imaging system can resolve. In the past decade numerous efforts have been made to achieve imaging resolution beyond that of Rayleigh limit. Among various proposals dedicated to surpass this "barrier", a representative example is the well-known technique of near-field imaging, an essential sub-wavelength imaging technology. Near-field imaging relies on the fact that the evanescent component carries fine details of the electromagnetic field distribution at the immediate vicinity of probed objects [1]. Over past decades, several innovative imaging instruments(e.g., the metamaterial superlens [2], metalens [3], etc.) have been invented, which detect the evanescent spectrum smartly to improve imaging resolution and operational distance of sensors to certain extent. Nonetheless, the strong confinement of measurements for near-field evanescent components precludes their widespread use as tools for general-purpose application. To circumvent this limitation, superoscillation based imaging has been developed as an alternative significant attempt for breaking Rayleigh limit [4]. Superoscillatory, laid first by Slepian et al, indicates that over a *finite* interval, a waveform oscillates arbitrarily faster than its highest component in its operational spectrum, and thus makes it possible to encode fine details of the probed object into the field of view beyond the evanescent fields. In light of this property, several optics devices have been built to achieve super-resolution imaging from far-field measurements [4, 5]. Although superoscillation-based imaging lifts the requirement of probe-object proximity, the obtainable enhancement in resolution is essentially dependent on SNR and others, and it requires a huge-size mask with enough fabrication finesse. In addition to the methods mentioned above, there are other approaches trying to beat the Rayleigh limit, for example, most recently Gazit et al proposed a technique of computational imaging, which is dedicated for sparse or compressible objects via solving a time-consuming nonlinear optimization problem constrained by sparse regularizer [6].

It is well known that the electromagnetic fields can carry not only energy and linear momentum but also angular momentum (decomposed into spin angular momentum, SAM, and orbital angular momentum, OAM) over very large distance [7]. It has been demonstrated that in the optics [8] and radio regimes [9], a beam with the distribution of helical phase front carries the information of OAM. Several solutions of generating this kind of vortex fields have been developed as well [9, 17]. OAM-carrying beams can provide more fruitful degrees of freedom for beam manipulation, and has been benefiting applications ranging from the information processing, communications [10, 11], to imaging mostly in the optical and quantum regimes [12]. However, less than ten years ago the researchers realized that the OAM held promises for resolving two optical sources separated at the distance lower than that of Rayleigh diffraction. Swartzlander discovered that the optical vortex mask with topological charge of 1 can be applied to distinguish two sources separated at the distance slightly lower than that of Rayleigh criterion [13]. Tamburini et al. found out that when two sources separated by an angular distance below the Rayleigh criterion crossed an optical vortex mask with topological charge of 1, the peak intensity ratio was highly sensitive to the distance between separated sources, hence force, can be used as a tool of resolving two sources spaced at the distance smaller than that of Rayleigh limit [14]. These methods aiming to break the Rayleigh limit seek to build heuristically the relation between measured OAM spectrum and some critical parameters of targets of interests, such as location, size and number of separated objects [15, 16]. In this letter, we revisit the problem of diffraction tomography (DT) and formulate a novel variant by exploiting the concept of OAM, termed as OAM-DT. The current study indicates that this new technology has no physical limits on imaging resolution in principle, and can be universally used for imaging of complicated objects.

Diffraction tomography is capable of not only retrieving quantitatively the distribution of dielectric constant of weak scattering objects, but also treating strong ones [18, 19]. To facilitate discussion, we adopt the configuration of transmission measurement, as sketched in Fig. 1. In this setup, there are two sets of coordinate systems, namely, the local system O-xyz is obtained by rotating the global system of O-XYZ around y-axis by $\theta$. The array of transmitters is distributed within the region denoted by a gray-filled square located at $z_t$ while the array of receivers at $z_r$. With the assumption of Born approximation, the OAM-carrying scattering field $E_{sca}(\mathbf{r},\ell)$ reads

$$E_{sca}(\mathbf{r},\ell) = \int G(\mathbf{r},\mathbf{r}') E_{in}(\mathbf{r}',\ell) O_{3D}(\mathbf{r}') d\mathbf{r}' \quad (1)$$

$\mathbf{r}=(x,y,z)\in D_r$, $\mathbf{r}'=(x',y',z')\in ROI$

Herein, $D_r$ is the region where receivers are placed, $O_{3D}(\mathbf{r}')$ represents the contrast of three-dimensional (3D) objects with respect to background medium (free space in this letter), $G(\mathbf{r},\mathbf{r}')=\dfrac{e^{jk_0|\mathbf{r}-\mathbf{r}'|}}{4\pi|\mathbf{r}-\mathbf{r}'|}$ is the Green's function in free space, and $k_0$ is the working wave number. Inside the region of interest (ROI) where probed objects are fallen into, the OAM-carrying incident wave $E_{in}(\mathbf{r}',\ell)$ with the topological charge of $\ell$ ($\ell=0,\pm1,\pm2,...,\pm L$) propagates along $z$ direction, i.e., $E_{in}(\mathbf{r}',\ell)=A(r')e^{j(\ell\varphi'+k_0 z')}$, where $\varphi'$ is the polar angular in the O-xy plane, and the core of this vortex field is along the z-axis. In our work $E_{in}(\mathbf{r}',\ell)=e^{j(\ell\varphi'+k_0 z')}$ is adopted. Bear in mind that only the use of $\ell=0$ corresponds to traditional DT technique.

In optics, numerous methods of generating OAM-carrying beams, such as with use of spiral phase plate (SPP), computer-aided hologram, etc..In radio, it was shown that an OAM-carrying radio beam can be generated by circular phased antenna array [9], or others [17]. In this letter the OAM-carrying field of $E_{in}(r')=A(r')e^{j(\ell\varphi'+k_0 z')}$ inside ROI is produced by a phase antenna array. In the simulate 400 elements of half-wave dipoles are uniformly distributed within the rectangle plane of [-5λ, 5λ]×[-5λ, 5λ] at $z_t$=-5λ. OAM-carrying radio beams can be generated by controlling adaptively phases and amplitudes of elements of the antenna array. In the top row of figure 2, the distributions of normalized amplitudes of resulting incident fields inside at the location of z=0 are illustrated for topological charges of $\ell$=0,1, 3, 5 and 10 from left to right in order. However, in the bottom row of figure 2, corresponding phase distributions are shown as well. Notice that the sizes of vortex fields follow the rule of $0.1|\ell|$, which means $E_{in}(r',\ell)\approx e^{j\ell\varphi'}$ for $r'\geq 0.1|\ell|$, and be approximately zero elsewhere.

Here we firstly investigate the ability of OAM-DT scheme for super-resolution imaging by examining the imaging of a thin dielectric slab, mathematically, $O_{3D}(\mathbf{r}') = O(x', y')\delta(z')$. For this case only one-view measurements are taken into consideration, specifically, $\theta = 0$. After applying the Weyl equation to the Green's function in Eq. (1), we obtain

$$E_{sca}(x,y,\ell) = \frac{j}{8\pi^2} \int_{-\infty}^{+\infty}\int_{-\infty}^{+\infty} dk_x dk_y \tilde{O}(k_x, k_y; \ell) \frac{e^{jk_x x + jk_y y + jk_z z_r}}{k_z} \qquad (2)$$

where $k_z = \sqrt{k_0^2 - k_x^2 - k_y^2}$, $\tilde{O}_\ell(k_x, k_y) = \int O(x', y') e^{j\ell\varphi'} e^{-jk_x x' - jk_y y'} dx' dy'$. Taking the 2D Fourier transform of $E_{sca}(x, y, \ell)$ with respect to $x$ and $y$ leads to

$$\tilde{E}_{sca}(k_x, k_y, \ell) \propto \frac{e^{jk_z z_r}}{k_z} (\tilde{O} \otimes \tilde{F}_\ell)(k_x, k_y) \qquad (3)$$

where $\otimes$ indicates convolution, $\tilde{O}(k_x, k_y) = \int O(x', y') e^{-jk_x x' - jk_y y'} dx' dy'$, and

$$\tilde{F}_\ell(k_x, k_y) = \int e^{-jk_x x' - jk_y y'} e^{j\ell\varphi'} r' dr' d\varphi' \propto j^{-|\ell|} e^{j\ell\theta_k} \int_0^R J_{|\ell|}(kr') r' dr' \qquad (4)$$

Here $\theta_k = \tan^{-1}\left(k_y/k_x\right)$. It is not hard to prove that in the far field of $k_0 z_r \gg 1$, $\tilde{E}_{sca}(k_x, k_y, \ell)$ is bandlimited to $k_x^2 + k_y^2 \leq k_0^2$ which is attributed to the Rayleigh limit for $\ell = 0$. In contrast, the convolution depicted by Eq. (3) with non-zero $\ell$ allows the conversion of high-frequency components of $\tilde{O}(k_x, k_y)$ outside $k_x^2 + k_y^2 \leq k_0^2$ into $\tilde{E}_{sca}(k_x, k_y, \ell)$ inside $k_x^2 + k_y^2 \leq k_0^2$, giving rise to the possibility of super-resolution imaging. It can be shown clearly by applying $\sum_{\ell=-\infty}^{\infty} e^{j\ell(\varphi-\varphi')} = \delta(\varphi - \varphi')$ to Eq.(3), the deduced imaging with arbitrarily high angular resolution, at least in principle, can be readily achieved. More strictly, we would like to look into this claim by considering so-called point-spread function (PSF) of OAM-DT. After some straightforward implementation, the expression of the reconstruction of $O(x', y')$ in terms of PSF, i.e.,

$$\hat{O}(x'', y'') = \int R(x', y'; x'', y'') O(x', y') dx' dy' \qquad (5)$$

$$(x', y'), (x'', y'') \in ROI$$

wherein the limit of $L \to \infty$ the PSF of OAM-DT is

$$R(x', y'; x'', y'') = \sum_{\ell=-\infty}^{\ell=+\infty} e^{-j\ell(\varphi'-\varphi'')} \int_{-\infty}^{+\infty}\int_{-\infty}^{+\infty} dk_x dk_y \frac{e^{-2\operatorname{Im}(k_z)z_r}}{|k_z|^2} e^{jk_x(x'-x'')+jk_y(y'-y'')}$$

$$= \delta(\varphi'-\varphi'') \int_0^{k_0} J_0(k|r'-r''|)k\,dk \quad (6)$$

$$= k_0 \delta(\varphi'-\varphi'') \frac{J_1(k_0|r'-r''|)}{|r'-r''|}$$

Eq. (6) reveals that with the use of topological charges varying from $-\infty$ through $\infty$, the extremely high angular resolution can be achieved while no obvious improvement on the radial resolution. Inspired by this observation, it is expected that imaging with super-resolution in the angular and radial directions can be obtained by changing cores of vortex incident fields.

Now, we consider the implementation of reconstructing a thin slab $O_{3D}(\mathbf{r}') = O(x', y')\delta(z')$ with the use of proposed OAM-DT technique. In matrix notation, Eq. (1) becomes

$$\mathbf{E}_\ell = \mathbf{A}_\ell \mathbf{O}_{3D} \quad (8)$$

where $\ell = 0, \pm 1, \pm 2, ..., \pm L$. Eq. (8) can be immediately solved via standard least square method; more specifically, this slab can be retrieved by the efficiently method solving quadrature optimization problem. Regarding the reconstruction of more general 3D objects, above discussions are applicable with multi-view measurements by changing $\theta$.

To illustrate previous investigations, a set of numerical experimentsare conducted, where the synthetic measurements are generated with the help of full-wave simulation of method of moment (MoM) [20] and corrupted by some additive Gaussian noise. In this simulation the region of interest (ROI) of 4λ×4λ×0.1λ, centered at the origin, is divided into 41×41×1 sub-blocks with size of 0.1λ×0.1λ×0.1λ, and only one-view measurements (i.e., $\theta = 0$) are served as the input of OAM-DT. The array of 30x30receiversare arranged uniformly within the rectangle region of [-5λ, 5λ]×[-5λ, 5λ] at the location of z=5λ, and distributed regularly within a square of 10λ×10λ with a spatial step of 0.5λ.As the first example, the target being imaged consists of a small cubic with size of

$0.1\lambda \times 0.1\lambda \times 0.1\lambda$ with refraction index of 1.01 located at $(0, \lambda, 0)$, and the SNR of measurements is set to be 40dB. Figure 3 compares reconstructions obtained by traditional DT and proposed OAM-DT technique. Fig.3 (a) gives the result by traditional DT method equivalent to the case of $L=0$, while Fig.3(b), (c), (d) and (e) are results by proposed OAM-DT with $L=3, 5, 10$ and 20. These results reveal that with the increase of truncation index $L$, reconstructed images become remarkably sharper, especially the strong enhancement on the angular resolution, as predicted previously. To improve the radial resolution further, we move the cores of OAM-carrying incident fields from $-1.5\lambda$ to $1.5\lambda$ with step of $0.5\lambda$ along the line of $x=\lambda$, and get the result as shown in Fig, 3(f). Here the image displayed an almost perfect reconstruction of "true" object. The second example is a more complicated imaging target which consists of three separated "U" structures with different orientations. In this simulation, all parameters are the same as above except that a lower SNR of 30dB. Figures 4 (b), (c) and (d) give, respectively, the results reconstructed by OAM-DT techniques with $L=0$ (traditional DT), 5, and 10. To improve imaging quality furthermore, cores of OAM-carrying incident fields are moved along the diagonal of ROI with step of $0.5\lambda$, and use $L=5$. The corresponding results demonstrated in Fig.4(e) reveal an almost perfect reconstruction of ground truth of Figure 4(a).

In the last example, we consider the reconstruction of 3D objects (as shown in Fig. 5(a)) by applying proposed OAM-DT technique, where simulation data are produced by using commercial software FEKO Suite 5.3 and then corrupted by white noise with SNR of 30 dB. In this case all parameters are the same as above. The simulation taking multi-view measurements of $\theta = 0°, 45°, 90°$ and $135°$. Fig.5 (c) shows the results of OAM-DT with $L=5$. The corresponding results computed by standard DT technique is displayed in Fig. 5(b) for comparison. Again, to further improve imaging quality, the centers of vortex fields are moved along the diagonal of ROI with step of 0.5, and the imaging results are shown in Fig. 5(d). A function of "*isofurface*" in Matlab 7.3 with threshold value of 0.0031is applied in the Fig. (c) and Fig. (d), while value of0.00078inFig. (b)due to relatively under-estimated values by traditional DT technique. For this example with 3D objects, conclusions drawn previously are verified again.

In summary, this letter demonstrates theoretically and numerically that it is indeed possible to significantly upgrade the imaging resolution of traditional diffraction tomography by applying electromagnetic OAM. This discovery serves a novel approach for sub-wavelength wavefield imaging in a rather efficient manner, which has no harsh requirements for elaborate imaging lens, near-field measurement, sophisticated and heavily computational post processing, and so on. It is conceivable that the proposed OAM-DT technique holds promise for the development of novel information-rich radar, smart antenna array, and other imaging systems. A concept-of-proof imaging system is under construction, and will be reported in the near further.

# Figures and Captions

FIGURE 1. The geometrical configuration adopted to study OAM-DT

FIGURE 2. The distributions of normalized amplitudes (top row) and phases (bottom) of OAM-carrying fields at z=0 produced by the phased antenna array located at $z_{t=}$-5$\lambda$. From left to right, they corresponds to $\ell$=0, 1, 3, 5, and 10 in order. In these figures, all axis are scaled by operational wavelength$\lambda$.

FIGURE 3.Reconstruction a small cubic with size of 0.1$\lambda$×0.1$\lambda$×0.1$\lambda$, and refraction index of 1.01. (a) L=0(equivalent to traditional DT method), (b)L=3, (c)L=5, (d)L=10, and (e)L=20. From Fig.(a) to Fig.(e), the cores of vortex field of incident fields are located at the origin of global coordinate system. Fig (f) is the result of moving the cores of incident vortex fields along $x$=$\lambda$. In all of these figures, all axis are scaled by operational wavelength $\lambda$.

FIGURE 4. (a) the ground truth. (b)the result obtained by traditional DT method. (c) and (d)the results of OAM-DT with L=5 and L=10, respectively. Fig.(e) the result of moving the cores of incident vortex fields along the diagonal of ROI. For all these figures, all axis are scaled by working wavelength$\lambda$.

FIGURE 5. (a) the ground truth. (b) the result obtained by traditional DT method. (c) the result of OAM-DT with $L$=5, where cores of incident vortex fields are fixed at the center of ROI . Fig.(d) the result of moving the cores of incident vortex fields along the diagonal of ROI. In these figures, all axis are scaled by working wavelength $\lambda$.

Figure 1

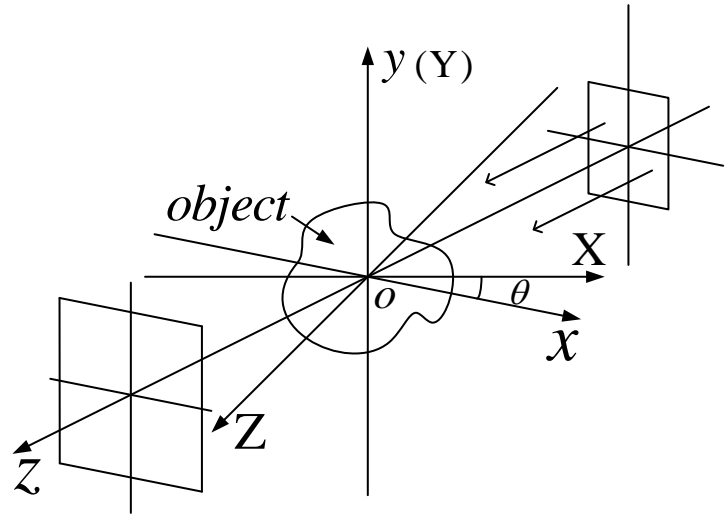

FIG.1 The geometrical configuration adopted to study OAM-DT

Figure 2

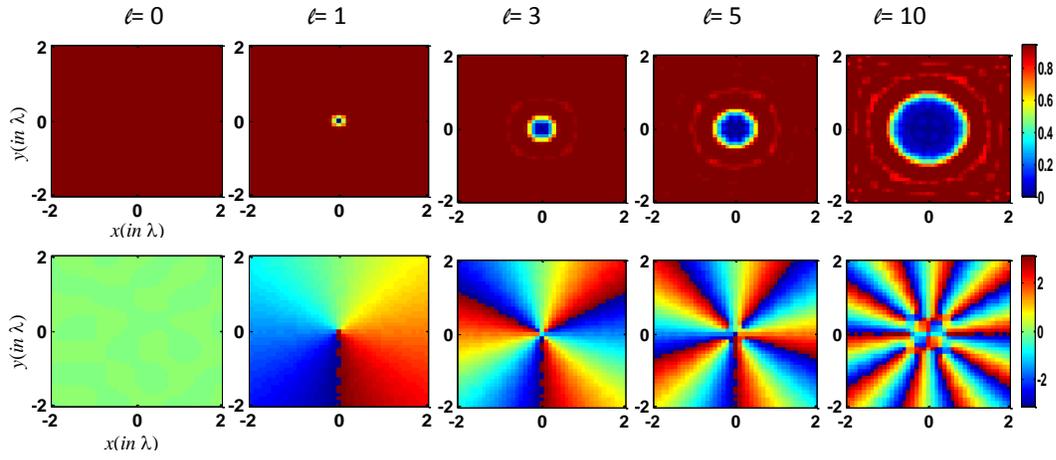

FIG. 2. The distributions of normalized amplitudes (top row) and phases (bottom) of OAM-carrying fields at z=0 produced by phased antenna array located at $z_{t=-5}\lambda$. From left to right, they corresponds to $\ell$=0, 1, 3, 5, and 10 in order. In these figures, all axis are scaled by operational wavelength λ.

Figure 3

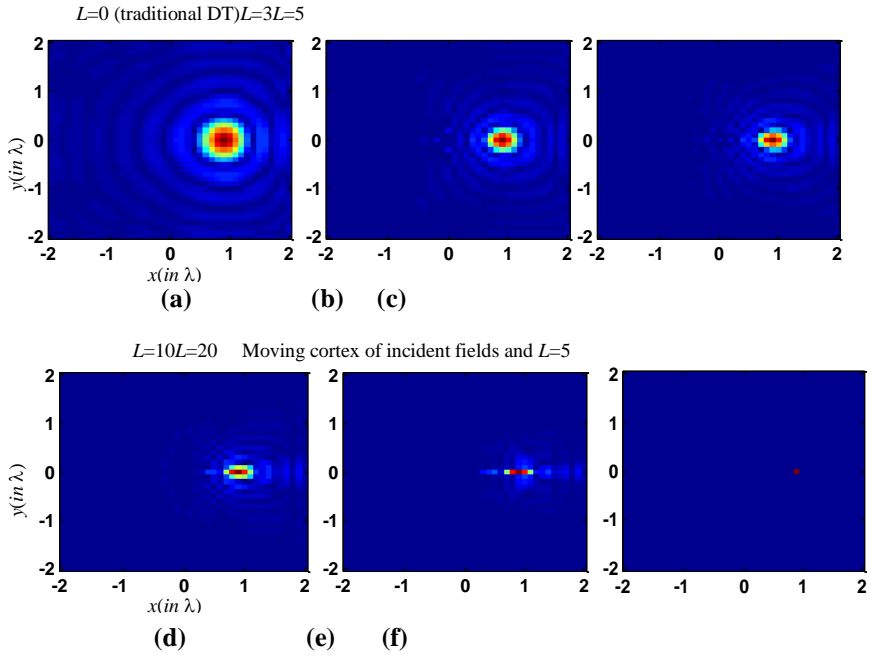

FIG. 3 Reconstruction a small cubic with size of 0.1λ×0.1λ×0.1λ, and refraction index of 1.01. (a) L=0(equivalent to traditional DT method), (b)L=3, (c)L=5, (d)L=10, and (e)L=20. From Fig.(a) to Fig.(e), the cores of vortex field of incident fields are located at the origin of global coordinate system. Fig (f) the result of moving the cores of incident vortex fields along $x=\lambda$. In all of these figures, all axis are scaled by working wavelength λ.

Figure 4

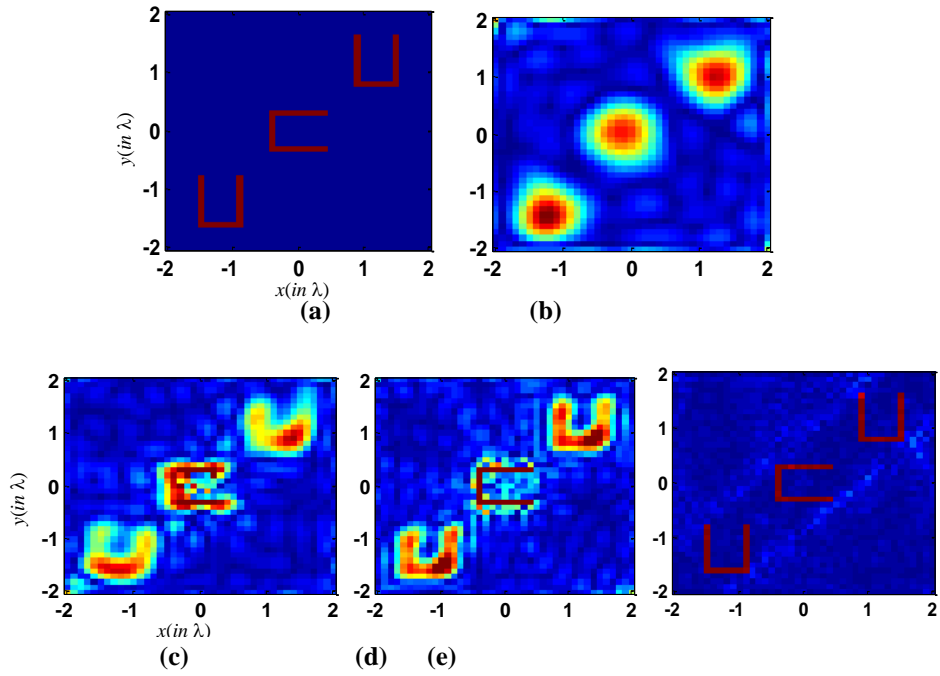

FIG. 4 (a) the ground truth. (b)the result obtained by traditional DT method. (c) and (d) the results of OAM-DT with L=5 and L=10, respectively. Fig.(e) the result of moving the cores of incident vortex fields along the diagonal of ROI. For all these figures, all axis are scaled by working wavelengthλ.

Figure 5

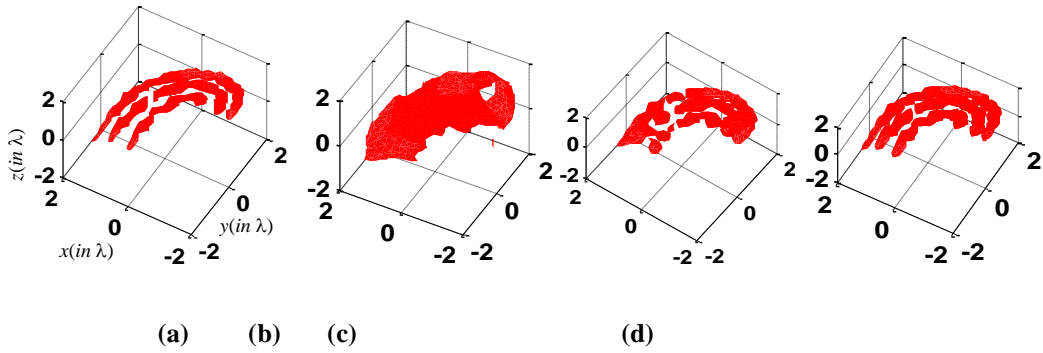

(a)   (b)   (c)           (d)

FIG. 5 (a) the ground truth. (b)the result obtained by traditional DT method. (c) the result of OAM-DT with *L*=5, wherecores of incident vortex fields are fixed at the center of ROI . Fig.(d) the result of moving the cores of incident vortex fields along the diagonal of ROI. In these figures, all axis are scaled by working wavelengthλ.